\documentclass[aps,pre]{revtex4-2}

\usepackage{graphicx}
\usepackage{amsmath}
\usepackage{amsfonts}
\usepackage{amssymb}

\begin{document}

\title{Modelling of impurity heating during reconnections in a Reverse Field Pinch device as due to parallel electric field acceleration and chaos-induced thermalization}

\author{F. Sattin}
\email{fabio.sattin@igi.cnr.it}
\address{
Consorzio RFX (CNR, ENEA, INFN, Universit\`a di Padova, Acciaierie Venete SpA), Padova, Italy}
\author{D.F. Escande}
\address{
 Aix-Marseille Universit\'e, CNRS, PIIM, UMR 7345, Marseille, France }
 \author{M. Gobbin, I. Predebon}
 \address{CNR ISTP, Istituto per la Scienza e Tecnologia dei Plasmi, Sede di Padova, Italy; Consorzio RFX (CNR, ENEA, INFN, Università di Padova, Acciaierie Venete SpA), Padova, Italy.}

\begin{abstract}
The ion temperature during magnetic reconnections measured along the direction of the magnetic field at the MST Reverse Field Pinch has not yet received a satisfactory theoretical explanation. In this  work we argue that it is consistent with a picture of ion energization by the parallel electric fields generated by the plasma during reconnection, and thermalization due to the chaotic ion dynamics. Three possible sources of randomness are pointed out: particle motion along static stochastic magnetic field lines, breakup of the adiabatic ion dynamics caused by the strong gradient nonlinearity of the magnetic field, parallel acceleration along stochastic time-dependent electric fields.
The three mechanisms are likely operative simultaneously; regardless the specific mechanism active, the correct scaling laws with ion parameters are recovered. Furthermore, we argue that quantitative agreement can be obtained by feeding the model with realistic values for the plasma conditions.  
\end{abstract}

\maketitle

\section{Introduction}
Reconnection of magnetic field lines is a basic process, ubiquitous wherever magnetized plasmas are present. During reconnection, a fraction of the energy accumulated in the magnetic field is delivered to the particle component of the plasma, ions and electrons. The ways by which energy is transferred are potentially several, a detailed assessment of the energy budget is an area of active investigation, both in astrophysical and laboratory plasmas; it is usually a difficult task, possible only in a few well-controlled environments (see, e.g., \cite{hsu00}); furthermore, it is likely that different mechanisms are active among different physical scenarios.  \\
The Reversed Field Pinch (RFP) is a toroidal device designed for fusion plasma research, characterized by a magnetic field structure that is mostly self-organized \cite{marrelli21}. The sustainment of quasi-steady-state conditions requires the generation of toroidal flux by the plasma itself, which occurs via continuous and quasiperiodic reconnection events. Accordingly, the physics of reconnection is a particularly important topic in these devices, from the theory as well as the experimental side  \cite{marrelli21,momo20,gobbin22}.\\
This work addresses the extensive campaigns of experiments carried out in the period 2008-2014 in the MST RFP device and summarized in a long list of works \cite{ganga08,tangri08,fiksel09,magee11,ren11,kumar13,cartolano14}. These papers provide a detailed characterization of both the particle and field components of the plasma during reconnections, including thermal and magnetic energy budget, time- and space-resolved measurements of the temperature along the directions both aligned and perpendicular to the magnetic field, scan over several ion species, frequency- and wavelength-resolved power spectra of magnetic fluctuations.  
The data allowed a comparison with the predictions from several of the most popular models of ion heating, as well as some models developed {\it ad-hoc}. None of them was found able to satisfactorily match all of the experimental evidence. Therefore, the mechanism behind ion heating in this kind of devices is so far unexplained.  
In paper \cite{kumar13} it was argued, in particular, that a complete explanation of the ion heating likely requires two separate mechanisms, delivering energy to the particles respectively along and perpendicularly to the magnetic field. \\
In this work we propose a plausible explanation for the findings of the MST team focussing on the interpretation of ion heating along the direction parallel to the magnetic field.  We show that it is quantitatively consistent with the predictions of a model of ion energization due to parallel electric fields coupled to some non-collisional thermalization mechanism. We identify three different possible origins for the electric field and the thermalization mechanism, likely co-existing.    
Very good agreement with the experiment may be obtained both in quantitative values and in terms of scaling laws as far as the ion energization along the direction parallel to the magnetic field is concerned. 
The structure of the paper is the following: next section provides a brief overview of the fundamental facts concerning ion heating in MST during reconnections; section III describes the model and suggests a fit of the experimental data on the basis of its predicted scaling. Some considerations about the reasonableness of the fitting coefficients are made. Concluding remarks are provided in section IV, where some open questions are also discussed, among them the issue of the energization perpendicular to the magnetic field, and the coupling between the parallel and perpendicular degrees of freedom. 

\section{A brief overview of the MST experimental results}
In this section, we summarize information collected by three papers.
First of all, Fiksel et al. \cite{fiksel09} monitored the behavior of plasma discharges with different main ions: $H, D, He$.  Measurements were performed using Rutheford scattering, which provides information about the energy distribution along a direction mainly perpendicular to the magnetic field. \\
Then, Kumar et al. \cite{kumar13} monitored the temperature of several charge states of different impurities with high time resolution in standard RFP Deuterium discharges using line-averaged spectroscopic methods.  Measurements refer to line emissions from edge impurities $Al^{1+,2+}, C^{2+,4+}, N^{2+,3+}, O^{2+,3+,4+}$  since the impact parameter of the lines of sight is about 90\% of the minor radius. Because of the experimental arrangement and of the radial position of the emitting species, the temperature measurement corresponds to the observation of Doppler broadening of spectral lines along a direction that is well aligned with the local direction of the magnetic field, hence the measurements refer to the parallel temperature.  \\
Thirdly, we recall the spectroscopic measurements of the parallel and perpendicular temperature of $C^{6+}$ impurity ions by Magee et al. which show comparable heating for both directions, with a mild excess of perpendicular heating \cite{magee11}. \\
We summarize the  main findings of these three works as follows: 
 (i) ion heating is extremely fast, it occurs simultaneously with reconnection over a time interval of roughly 50 $\mu$s or less. For comparison, the cooling phase is slower, by a threefold factor at least. (ii) Expressed in terms of power, the heating phase entails an absorption of a few MeV/s per ion. (iii) The final temperature increment ranges between a few tens to some hundreds of eV's. (iv) For one and the same ion, the levels of heating in the direction parallel and perpendicular to $\bf B$ are comparable, with a moderate predominance of perpendicular heating. (v) The scalings with the ion properties are quite different, though. Perpendicular heating is strongly correlated with the ion mass $M$: the increase in temperature scales roughly with $M^{0.5}$ whereas no appreciable dependence from the ion charge $Q$ is found.  Conversely, parallel heating depends on both $Q$ and $M$ following a scaling like $Q^\gamma/M$, with $ \gamma$ close to the unity.  

\section{Modelling of the parallel ion heating } 
Any interpretation of the experimental data must account for the list of results (i-v). \\
The full heating process may be modeled formally through two coupled rate equations for the parallel and perpendicular temperatures $T_{\parallel, \perp}$ (see Eqns. 1,2 of \cite{kumar13}) which include separate heat sources along the two directions, isotropization terms which mix the parallel and perpendicular direction, and loss terms towards the background plasma. Kumar et al. made a  qualitative study of the equations and showed that including just one heat source is sufficient to raise the temperature along both directions, thanks to the isotropization terms. The correct scaling along one direction may be achieved, provided that the source has the required functional dependence upon ion parameters, but not along the two directions simultaneously. \\
Conceptually, the present work is an advancement of the study done by Kumar et al. Consistently with their claims, we drop any attempt of describing the heating process in its wholeness, and limit to consider heating along the parallel direction alone. Specifically, we will address, out of the previous list, only the points (i,ii,iii) and partially (v). The qualifying aspect of the work is the choice of the terms to be fed into the rate equation: the energy source and the isotropization mechanism. Concerning the first, 
it is well known that electric fields parallel to the magnetic one develop during reconnections \cite{egedal12,ergun16,andersonaip16,anderson16}, thus one natural mechanism for particle energization is provided by acceleration along these fields. 
Some thermalization mechanism must be invoked as well, since otherwise all identical particles would be accelerated by the same amount and this does not produce a broad energy distribution. Mechanisms based upon binary collisions seem too slow to account for the measurements. For instance,  according to ref. \cite{NRL} the collision frequency of a $C^{4+}$ ion (which is a best-case example, given its high charge--to--mass ratio) in a $O(100) \, eV, O(10^{19}) \, m^{-3}$ Deuterium background is few tens of kHz whereas, to be compliant with the experiment, frequencies close to 100 kHz are needed. Indeed, the very Figure 5 of ref. \cite{kumar13}, produced using collisional estimates, shows plasma dynamics evolving over time scales slower than the experimental ones. The discrepancy is not of orders of magnitude, yet is large enough to be noticeable, hence a non-collisional thermalization mechanism must be invoked. 
In this work we will review three potential candidate mechanisms. In the first one, thermalization is provided by some weakly chaotic stationary magnetic field.We argue that this mechanism could be the most likely operative one, since a lot of magnetic chaos is actually identified during reconnections, see Fig. 10 in \cite{gobbin22}. The second mechanism is popular mostly among the astrophysics community.  
It has been recognized for a long time that in the neighborhood of the magnetic neutral point or, more generally, of a local minimum of a spatially inhomogeneous magnetic field, even though the field itself is smooth, the orbital motion of charged particles may become chaotic due to the strong gradient nonlinearity \cite{garren58,hastie68,howard71,foote72,timofeev78,martin93} as soon as the ratio between the particle Larmor radius and the typical length scale for the variation of $B$ takes locally values of the order of unity. 
The two mechanisms are not necessarily alternative, and may coexist: recently, Lemoine \cite{lemoine23} argued that fully developed MHD turbulence produces a spectrum of magnetic structures whose sizes span all the way down to the ion Larmor radius. It produces simultaneously, therefore, both magnetic chaos and breakup of the particle's adiabatic dynamics.  
Thirdly, we acknowledge that reconnections are actually dynamic process, involving a broadband spectrum of magnetic activity. Fluctuating electric fields--with a component parallel to the equilibrium magnetic field--are thus inductively generated.    

\subsection{Particle heating in chaotic static magnetic fields} 
\label{subsection1}
Let us write the magnetic and electric fields as ${\bf B } = (b_x, 0, B), {\bf E} = (0,0,E)$. Both the guide field $B$ and the electric field $E$ are taken constant in space and time, whereas $ |b_x| \ll B$ is chaotically varying in space. \\
 The projection of the particle equation of motion along $z$ is
 \begin{equation}
 {  d v_z \over dt} = {Q \over M} \left( E - b_x v_y \right)
 \end{equation}
 and the drift along $y$ is $v_y =  E b_x/B$, thus
 \begin{equation}
   {  d v_z \over dt} = {Q \over M} E  \left( 1 - {b_x^2 \over B^2} \right) \to v_z = v_z(0) +   {Q \over M} E  \int dt  \left( 1 - {b_x^2 \over B^2} \right) 
 \label{eq:dvz}  
\end{equation}
In the previous equation the quantity $  f =  \int dt  \left( 1 - (b_x/B)^2 \right) $ appears. It is a stochastic variable, since the term $b_x = b_x(y(t),z(t))$ that appears inside the integral depends on the trajectory. \\ 
The temperature of an ensemble of particles is proportional to the variance of the velocity distribution:
 \begin{equation}
 T = {1 \over 2} M \langle (v_z -\langle v_z\rangle)^2 \rangle =  {1 \over 2} M \left( \langle  v_z^2 \rangle - \langle v_z \rangle^2 \right)
 \end{equation}
 Eventually, the temperature increment is (assuming $\langle v_z(0) \rangle = 0$):
 \begin{equation}
 \Delta T = T(t) - T(0) =  { Q^2 \over 2 M} E^2 \left( \langle f^2 \rangle - \langle  f \rangle^2  \right)
 \label{eq:dk}
\end{equation}
If $b_x$ were approximately constant along the particle motion, $ \langle f^2 \rangle \approx \langle  f \rangle^2 $, and no heating would take place. \\
In order to assess whether Eq. \ref{eq:dk} may be quantitatively consistent with the data, we write $ Q = Z \, e,   M  =\mu \, m_p, \left( \langle f^2 \rangle - \langle  f \rangle^2  \right) = k^2 t^2$, with $e, m_p$ the proton charge and mass and $k <1$. If we replace for $t$ the typical duration  of reconnection, say $ t  = 5\times10^{-5}$s, we obtain
\begin{equation}
\Delta T \, (eV) ={Z^2 \over \mu}   0.25  \times(E k)^2
\end{equation}
requiring $E k$  to be at least $O(10^2)$  V/m in order to get $\Delta T \approx O(10) eV$  (the ratio $Z^2/\mu$ varies in the range $ \equiv 0.03 \div 1$ for the ions studied in MST). The estimates we have for for $E$ vary from $ O(10^2)$ V/m (see Fig. 2b  in  \cite{anderson16}) to $O(10^3)$ V/m \cite{fiksel09}.  
  
 \subsection{Heating in static electric fields and non-uniform smooth magnetic fields}
 \label{subsection2}
In the neighborhood of a local minimum of a spatially inhomogeneous magnetic field, the orbital motion of charged particles may become chaotic when the ratio between the particle Larmor radius and the typical length scale for the variation of $B$ takes locally values of the order of unity. This result was employed to justify particle heating by very-low-frequency waves \cite{escande19}, poor confinement of fast ions in spherical tokamaks \cite{carlsson01,yavorskij02,escande21} or of charged particles in the magnetosphere \cite{martin86,buchner89,young02,liu22}. Mason and Rusbridge \cite{mason79} and, later on, Yoshida et al. \cite{yoshida98,numata02,numata03}, pointed out that the chaotic motion of charged particles may be equivalent to the randomization process of the directional motion of the current-carrying particles induced by an effective collisionality. Numerical simulations \cite{wang21} show that this anomalous collisionality is much larger than the Spitzer one by several orders of magnitude in low-density astrophysical plasmas, and easily by a ten- to hundredfold factor in laboratory ones, potentially providing the amplitude required. \\
Formally, all results are an exact replica of the previous subsection: Chaos mixes parallel and perpendicular motion; along the direction aligned with the electric field (say, $z$) particles still experience an accelerated motion but with an effective average smaller acceleration, just like  in Eq.\ref{eq:dvz}:
\begin{equation}
v_z \approx {Q \over M} \alpha E \, t
\label{eq:1}
\end{equation}
Here, we have employed $\alpha < 1$ rather than $k$ just to emphasize its different physical origin. 
Furthermore, the motion along the perpendicular direction becomes diffusive \cite{shang17}. 

\subsection{Heating in fluctuating electric fields}
\label{subsection3}
Magnetic activity has a broadband distribution both in frequency (see fig. 1 in \cite{ren11} and fig. 4 in \cite{kumar13}) and wavenumbers (fig. 2 in \cite{ren11}). Let us consider the two model vector potentials
\begin{equation} 
{\bf A}_0 = B_0 x  \hat{y} , \quad {\bf A}_1 =  b_w/k \cos (k x - \omega t) \hat{z}
\end{equation} 
The magnetic fields ${\bf B}_0 = \nabla \times {\bf A}_0,   {\bf B}_1 = \nabla \times {\bf A}_1$ stand respectively for the homogeneous equilibrium field, and a generic Fourier component picked up from the whole magnetic spectrum.  The term ${\bf A}_1$ stands for a transversal magnetic wave propagating along a direction perpendicular to the equilibrium field. It does not cause loss of generality; furthermore, it is consistent with the experimental fact that the spectral density is centered close to $k_\parallel \approx 0$.   
An electric field ${\bf E}_1$ is produced
\begin{equation}
{\bf E}_1  = - {\partial {\bf A}_1 \over \partial t} = - { \omega b_w \over k} \sin ( k x - \omega  t ) \hat{z}
\end{equation}
Hence, the particle experiences an acceleration along the parallel direction, effectively aligned with the z-axis since we postulate $b_w \ll B_0$:
\begin{equation}
M {dv_z \over dt} \approx Q E_1   \to v_z(t) = v_z(0) -  {Q \over M}  {\omega b_w \over k} \int dt  \sin (k x - \omega t) 
\end{equation} 
The argument then proceeds like in subsection \ref{subsection1}, if one supposes that the term $\sin (k x - \omega t) $ behaves like a stochastic variable, since $x$ depends on the specific trajectory \cite{dasgupta14}. This is likely for a full turbulent spectrum of magnetic fluctuations--which is found experimentally. We remark, however, that in principle chaos may take place even with a single wave. 
  
\subsection{Fitting of the experimental data}
\begin{figure}[h]
\includegraphics[width=8cm]{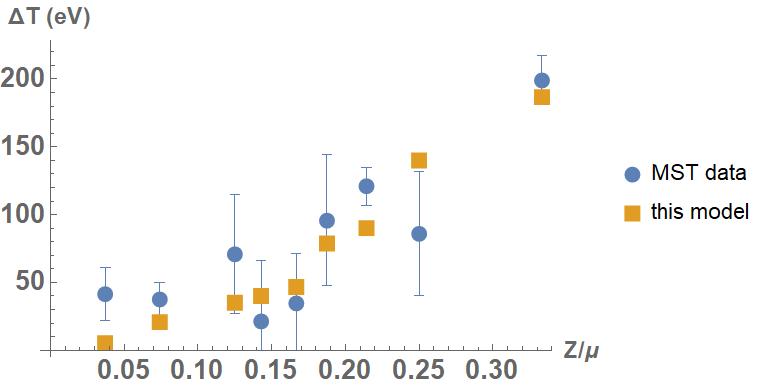}
\caption{Parallel temperature increase {\it versus} $Z/\mu$. Blue circles with error bars: data acquired from Fig. 3 of \cite{kumar13}; orange squares, best fit using the present model (Eq. \ref{eq:4}), with $c = 140$ eV. }
\label{fig:1}
\end{figure}

In conclusion, regardless of the precise mixture of mechanisms active we expect a  scaling like 
\begin{equation}
\Delta T = c \, (Z^2/\mu) \; eV. 
\label{eq:4}
\end{equation}
This scaling is generic, and arises whenever there is acceleration by given fields. \\
In this subsection we look at the original data presented in \cite{kumar13} and argue that: (i) even if the discussion therein suggested $Z /\mu$ as scaling, the present $Z^2 /\mu$ proposal performs at least as well, when employed to fit the data. (ii) The absolute values matter, too. In order to match experiments, we expect $ c \approx O(10^2) eV$, and we have argued in subsection \ref{subsection1} that this estimate does not seem unrealistic on the basis of the known plasma conditions, at least for the first mechanism proposed.  \\
We digitally acquired the data from Fig. 3 of \cite{kumar13} and made a best fit using Eq.\ref{eq:4} . The result is shown in Fig. \ref{fig:1} in terms of the original scaling $Z/\mu$.  The agreement appears excellent, the fitting parameter is $c = 140 $ eV which perfectly matches our order-of-magnitude estimates.            
  
\section{Summary and concluding considerations}
We have shown that a model of particle energization by parallel electric fields coupled with chaos-induced thermalization is compatible with MST data of impurity parallel heating during reconnections. We obtained satisfactory results both in terms of absolute values and in terms of scaling relations. The model involves just a few free parameters, whose best-fit numerical values turn out to be fairly reasonable. \\
The difference between our modeling and Kumar's is in the choice of the ingredients: the thermalization mechanism and the energy source. We replace their collisional thermalization with the chaos-induced one since argue that the former cannot quantitatively account for the data. Furthermore, the existence of parallel electric fields is quite well established, hence it appears natural to take them into account. In the present work, we have considered just the parallel heating source, nonetheless, it is unavoidable that chaos does not just thermalize the parallel velocity distribution, it also couples parallel motion with the perpendicular one. 
This entails that a consistent interpretation of ion heating must involve a comprehensive picture, where both parallel and perpendicular heating are included and treated simultaneously, with feedback from one term to the other. This is not a simple exercise, which we do not attempt to carry out here, and is further complicated by the fact, as we have argued earlier, that presently no candidate mechanism for perpendicular heating is fully convincing. Stochastic heating by low-frequency electromagnetic turbulence might be a promising candidate, except that the level of turbulence in RFP devices is definitely too small, even though some advancements have recently been made \cite{sattin23}.

\section*{AUTHOR DECLARATIONS}

\subsection*{Conflict of Interest}
The authors have no conflicts to disclose.

\subsection*{Author Contributions}
F. Sattin: Conceptualization (lead); Formal analysis (lead);
Investigation (lead); Validation (lead); Writing – original draft
(lead); Writing – review \& editing (equal). 
D.F. Escande, I. Predebon, M. Gobbin:
Conceptualization (supporting); Formal analysis (supporting);
Investigation (supporting); Validation (supporting); Writing – review \& editing (equal). 

\subsection*{DATA AVAILABILITY}
The data that support the findings of this study are available
within the article and the articles listed in the references section below

\end{document}